\documentclass[letterpaper]{article} 
\usepackage{aaai2026}  
\usepackage{times}  
\usepackage{helvet}  
\usepackage{courier}  
\usepackage[hyphens]{url}  
\usepackage{graphicx} 
\usepackage{natbib}  
\usepackage{caption} 
\usepackage{algorithm}
\usepackage{algorithmic}

\usepackage{bm}
\usepackage{amssymb}
\usepackage{amsmath}
\usepackage{array}
\usepackage{booktabs}
\usepackage{subcaption}
\usepackage{multirow}
\usepackage{newfloat}
\usepackage{listings}
\DeclareCaptionStyle{ruled}{labelfont=normalfont,labelsep=colon,strut=off} 
\floatstyle{ruled}
\newfloat{listing}{tb}{lst}{}
\floatname{listing}{Listing}
\title{Sonic4D: Spatial Audio Generation for Immersive 4D Scene Exploration}

\author {
    Siyi Xie\textsuperscript{\rm 1}\footnotemark[1],
    Hanxin Zhu\textsuperscript{\rm 1,\rm 2}\footnotemark[1]\footnotemark[2],
    Xinyi Chen\textsuperscript{\rm 1},
    Tianyu He\textsuperscript{\rm 3},
    Xin Li\textsuperscript{\rm 1}\footnotemark[3],
    Zhibo Chen\textsuperscript{\rm 1,\rm 2}\footnotemark[3],
}
\affiliations {
    \textsuperscript{\rm 1}University of Science and Technology of China\\
    \textsuperscript{\rm 2}Zhongguancun Academy, Beijing, China\\
    \textsuperscript{\rm 3}Microsoft Research Asia\\
    \{ustc2020xsy, hanxinzhu, chenxinyi0022\}@mail.ustc.edu.cn,\\ 
    tianyuhe@microsoft.com , \{xin.li, chenzhibo\}@ustc.edu.cn\\
    \texttt{https://x-drunker.github.io/Sonic4D-project-page/}
}

\usepackage{bibentry}

\begin{document}

\maketitle
\renewcommand{\thefootnote}{\fnsymbol{footnote}}
\footnotetext[1]{Equal Contribution}
\footnotetext[2]{Project Lead}
\footnotetext[3]{Corresponding Authors}

\begin{abstract}
Recent advancements in 4D generation have demonstrated its remarkable capability in synthesizing photorealistic renderings of dynamic 3D scenes. However, despite achieving impressive visual performance, almost all existing methods overlook the generation of spatial audio aligned with the corresponding 4D scenes, posing a significant limitation to truly immersive audiovisual experiences. To mitigate this issue, we propose \textbf{Sonic4D}, a novel framework that enables spatial audio generation for immersive exploration of 4D scenes. Specifically, 
our method is composed of three stages: 1) To capture both the dynamic visual content and raw auditory information from a monocular video, we first employ pre-trained expert models to generate the 4D scene and its corresponding monaural audio. 2) Subsequently, to transform the monaural audio into spatial audio, we localize and track the sound sources within the 4D scene, where their 3D spatial coordinates at different timestamps are estimated via a pixel-level visual grounding strategy. 3) Based on the estimated sound source locations, we further synthesize plausible spatial audio that varies across different viewpoints and timestamps using physics-based simulation. Extensive experiments have demonstrated that our proposed method generates realistic spatial audio consistent with the synthesized 4D scene in a training-free manner, significantly enhancing the immersive experience for users. 
\end{abstract}



\section{Introduction}
Benefiting from large-scale data available~\cite{soomro2012ucf101,schuhmann2021laion,yu2023celebv} and recent advancement in generative models~\cite{ho2020denoising,song2020denoising}, 4D generation (\textit{i.e.}, dynamic 3D scene generation)~\cite{jiang2024consistentd,ren2023dreamgaussian4d,zeng2024stag4d,zhu2025ar4d} has emerged as a promising direction due to its powerful capability in modeling complex spatiotemporal dynamics of real-world scenes. By enabling spatiotemporally consistent renderings from arbitrary camera viewpoints, 4D generation facilitates various downstream applications such as AR/VR~\cite{li20244k4dgen,fritsch20173d}, robotics~\cite{khalid20224d,hann20204d}, and autonomous driving~\cite{wang2024occsora,min2024driveworld}.

However, while existing 4D generation methods~\cite{hann20204d,liu2025free4d,zhang20244diffusion,zeng2024trans4d} achieve impressive visual results, they typically neglect the generation of spatial audio consistent with the corresponding 4D scene (\textit{i.e.}, audio that varies with the listener's viewpoint and follows physical acoustic principles), limiting the overall immersive experience to a large extent.

\begin{figure*}[t]
    \centering
    \includegraphics[width=1\linewidth]{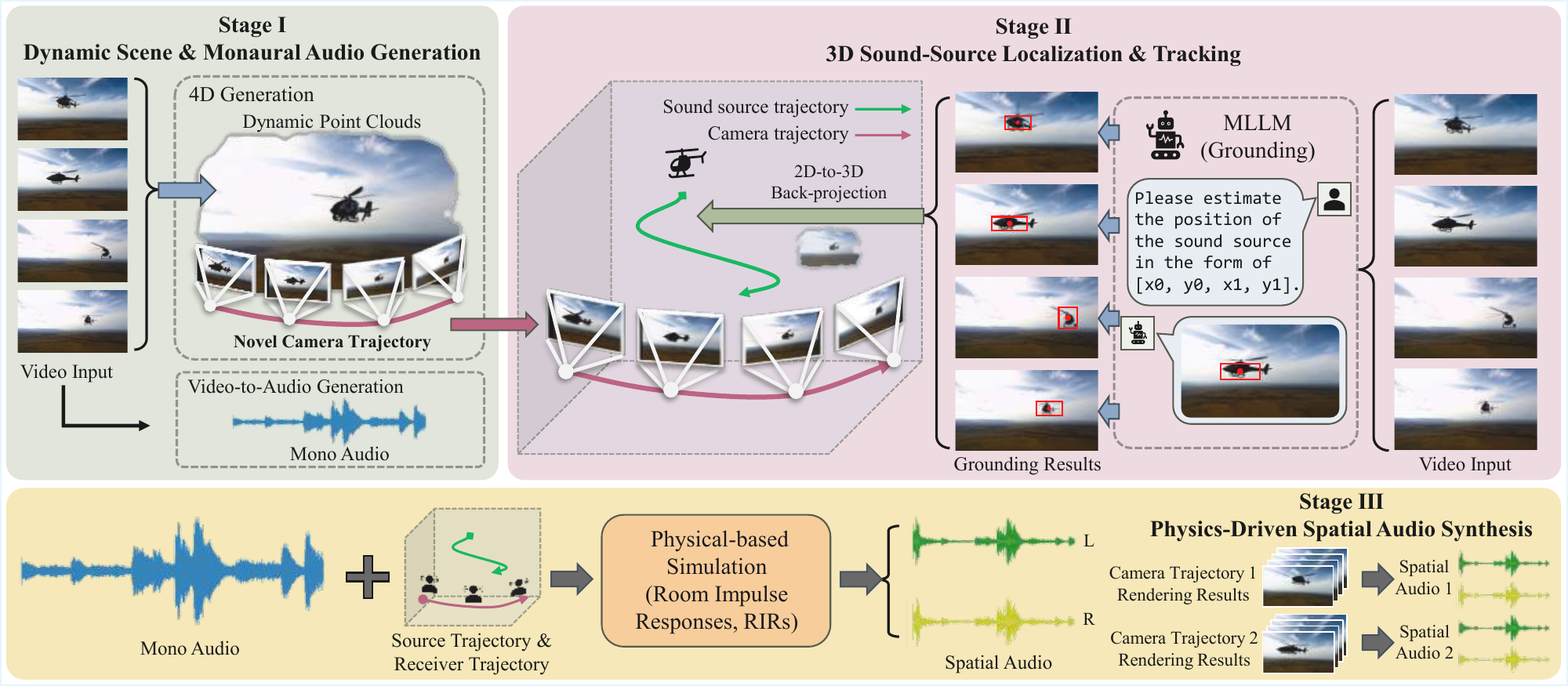}
    \caption{Pipeline of Sonic4D. Our method is composed of three stages: 1) Dynamic Scene and Monaural Audio Generation: extracting semantically aligned visual scenes and audio priors from monocular videos; 2) 3D Sound-Source Localization and Tracking: estimating the sound source's trajectory in 3D space for physically accurate sound propagation modeling; 3) Physics-Driven Spatial Audio Synthesis: leveraging a physics-based room impulse response simulation to realize spatial audio simulation.}
    \label{fig:pipeline}
\end{figure*}

To address this limitation, in this paper we propose Sonic4D, a novel framework that enables free-viewpoint rendering of both dynamic visual content and spatially consistent audio, thereby ensuring more immersive audiovisual exploration within the generated 4D scene. To this end, as shown in Fig.~\ref{fig:pipeline}, we design a three-stage pipeline: \textbf{1)} \textbf{Dynamic Scene and Monaural Audio Generation.} To empower novel-view visual renderings and provide essential spatial priors for spatial audio generation, we first leverage a pre-trained 4D generative model to synthesize the dynamic 3D scene from a monocular video. In parallel, a video-to-audio generative model is utilized to produce monaural audio that is semantically aligned with the generated 4D scene, serving as the raw acoustic input for subsequent spatial audio rendering. \textbf{2)} \textbf{3D Sound-Source Localization and Tracking.} To enable accurate physical simulation of dynamic sound propagation, we further estimate the sound source's trajectory in 3D space (\textit{i.e.}, the 3D locations of the sound source at different timestamps). To achieve this goal, we first use a multimodal large language model (MLLM) to perform pixel-level visual grounding on the input video, identifying the 2D coordinates of the sound source in each frame. These 2D positions are then back-projected onto the dynamic point cloud reconstructed in the previous stage, yielding a sequence of 3D coordinates that represent the sound source's trajectory. \textbf{3)} \textbf{Physics-Driven Spatial Audio Synthesis.} 
Given the estimated sound source location and the generated 4D environment, a physics-based simulation using acoustic room impulse responses (RIRs) is employed to simulate plausible spatial audio, enabling more immersive experiences in complex 4D scenes.

Notably, our method adopts a modular architecture that facilitates the integration of state-of-the-art pre-trained expert models, ensuring that the framework remains extensible and future-proof, with performance continually improving as more advanced expert models become available.

The main contributions of this paper can be summarized as follows:
\begin{itemize}
    \item We propose Sonic4D, a novel framework that achieves spatial audio generation for immersive 4D scene exploration. To the best of our knowledge, this is \textbf{the first work} that introduces spatial audio into the context of 4D generation.
    \item We propose a three-stage framework to achieve spatial audio generation: \textbf{1) Dynamic Scene and Monaural Audio Generation} to extract semantically aligned visual and audio priors from a single video; \textbf{2) 3D Sound-Source Localization and Tracking} to recover sound source's 3D trajectory for precise acoustic simulation; \textbf{3) Physics-Driven Spatial Audio Synthesis} to render dynamic, viewpoint‑adaptive binaural audio via physics-based room impulse response simulation.
    \item Extensive experiments have demonstrated  that our framework can effectively generate spatial audio consistent with 4D visual content, enabling much more immersive and coherent audiovisual experiences.
\end{itemize}


\section{Related Work}

\subsection{4D Generation}Early advances in 4D generation have built upon techniques such as Score Distillation Sampling (SDS) \cite{poole2022sds}, which leverages generative priors of pre-trained diffusion models to optimize dynamic scene representations. This paradigm laid the groundwork for methods such as Dream-in-4D~\cite{zheng2024unified}, which further integrated motion priors from pretrained video diffusion models into dynamic NeRF-based representations~\cite{pumarola2021d,yan2023nerf}. Concurrently, models like Consistent4D~\cite{jiang2024consistentd} and 4Diffusion~\cite{zhang20244diffusion} focused on enhancing spatiotemporal consistency, utilizing multiview interpolation, temporal alignment modules, or synchronized training strategies. Works such as Diffusion4D~\cite{liang2024diffusion4d}, PLA4D~\cite{miao2024pla4d}, and 4Dynamic~\cite{yuan20244dynamic} further emphasized controllability and efficiency by leveraging pixel-aligned supervision, video-based guidance, or hybrid representations combining mesh and Gaussian structures. Recent studies such as Free4D~\cite{liu2025free4d} and MVTokenFlow~\cite{huang2025mvtokenflow} further improve spatial-temporal coherence and visual quality without tuning large generative models. 

Though achieving remarkable results, all these methods focus solely on visual content rendering, neglecting the generation of spatial audio that aligns with the dynamic scene. In contrast, in this paper we propose a novel framework for spatial audio generation that enables more coherent and immersive exploration of 4D scenes from arbitrary viewpoints.

\subsection{Spatial Audio Generation}

Early spatial audio generation methods ~\cite{gao20192, xu2021visually, leng2022binauralgrad} predominantly took monaural inputs and “binauralized” them via neural networks conditioned on visual cues. In recent years, the advent of powerful generative architectures has spurred end‑to‑end models that accept text prompts~\cite{sun2024both}, images~\cite{sun2024both, dagli2024see}, spatial parameters~\cite{heydari2025immersediffusion,kushwaha2025diff}, silent videos~\cite{kim2025visage}, or full 360° panoramas~\cite{liu2025omniaudio} to jointly learn semantic and spatial audio features. For example, SpatialSonic~\cite{sun2024both} leverages spatial‑aware encoders and azimuth state matrices within a latent diffusion framework to provide fine‑grained multimodal spatial guidance for stereo audio generation. ViSAGe~\cite{kim2025visage} fuses CLIP visual embeddings with autoregressive neural audio codec modeling to generate coherent FOA directly from silent video frames. Recently, OmniAudio~\cite{liu2025omniaudio} addressed the 360° spatial audio gap by using a dual‑branch architecture to fuse panoramic and FoV streams, enabling high‑fidelity FOA generation from full spherical video content.

Although these generative methods achieve impressive multimodal results, they still face two key limitations. First, data scarcity and domain gaps in end-to-end learning restrict the fidelity of spatial audio rendering. Second, even with 360° video inputs, the synthesized sound field is bound to a fixed camera pose and cannot support dynamic listener motion or free-viewpoint changes. In contrast, our method decouples semantic audio synthesis from spatial rendering and leverages RIR-based convolution to physically simulate spatial acoustics for a moving listener in dynamic scenes.

\subsection{Sound Source Localization}

Sound source localization aims to find the location of sound sources in an image or video frame. Initial studies~\cite{senocak2018learning, chen2021localizing, senocak2023sound, um2023audio} addressed this problem by leveraging methods like cross-modal attention and contrastive learning to establish effective alignment between audio and visual modalities, mainly focusing on single sound source localization. Some works have taken this further to achieve multi-sound source localization~\cite{mo2023audio, kim2024learning}, enabling the simultaneous localization of multiple sound sources from mixed audio and visual inputs. However, these approaches focus on exploring audio–visual alignments and cannot localize sound sources without ground-truth audio. Recently, the rapid advancement of Multimodal Large Language Models~\cite{munasinghe2024videoglamm,li2024groundinggpt} (MLLMs) has made it possible to perform audio-visual grounding with strong generalization capabilities. In this work, we leverage the powerful prior knowledge of pretrained MLLMs to directly perform pixel-level visual grounding on the video input for sound source localization, followed by back-projection into 3D space to achieve accurate 3D sound source positioning and further spatial audio simulation.

\begin{figure*}[t]
    \centering
    \includegraphics[width=1\linewidth]{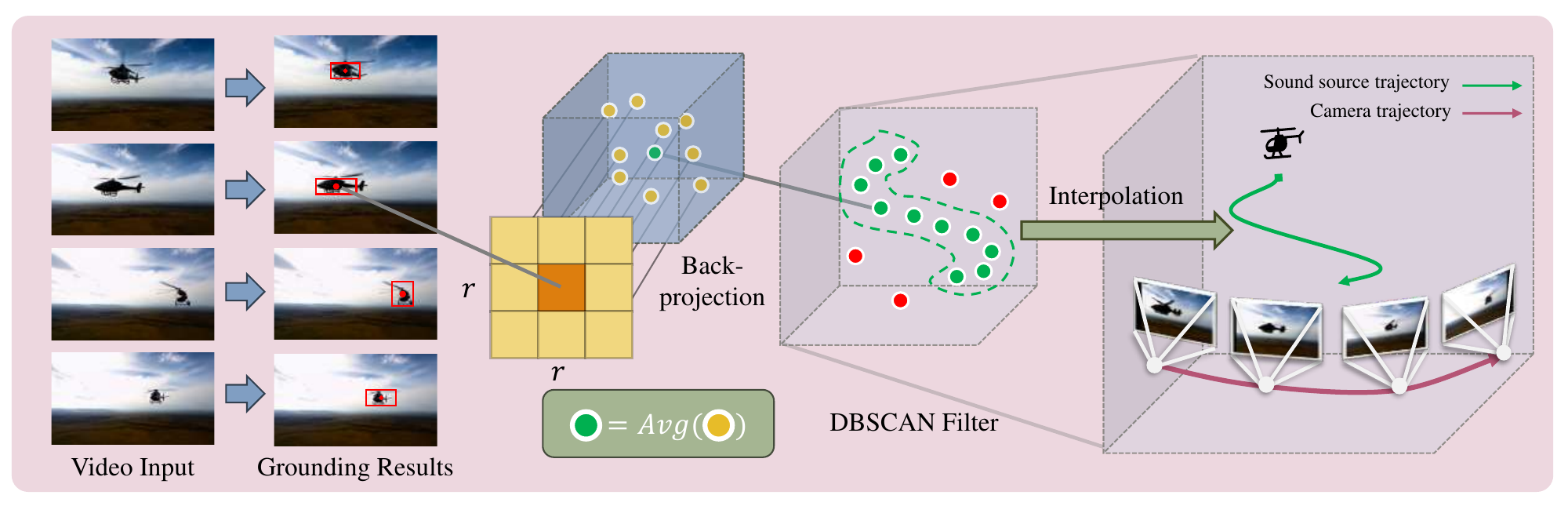}
    \caption{Illustration of Stage \uppercase\expandafter{\romannumeral2}: We localize the sound source in each frame using GroundingGPT~\cite{li2024groundinggpt}, back-project the 2D grounding results to 3D via depth, and apply DBSCAN~\cite{ester1996density} to obtain a smooth trajectory.}
    \label{fig:stage2}
\end{figure*}
\section{Methods}
Given a monocular input video $\bm{V}^s=\{\bm{I}^s_i\}_{i=1}^n\in \mathbb{R}^{n \times 3 \times H \times W}$, our goal is to generate both novel-view videos and corresponding spatial audio given arbitrary target camera trajectory $\bm{T}^r=\{\bm{T}_i^r\}_{i=1}^n\in \mathbb{R}^{n \times 4 \times 4}$, where each matrix $\bm{T}_i^r$ denotes the homogeneous camera pose at time $i$. As shown in Fig.~\ref{fig:pipeline}, the overall pipeline is divided into three stages: 1) Dynamic scene and monaural audio generation; 2) 3D sound-source localization and tracking; 3) Physics-driven spatial audio synthesis. Each stage is discussed in detail in the following subsections.

\subsection{Stage \uppercase\expandafter{\romannumeral1}: Dynamic Scene and Monaural Audio Generation}\label{Dynamic Scene and Monaural Audio Generation}
To obtain a novel-view video $\bm{V}^r$ while simultaneously capturing dynamic spatial priors for spatial audio simulation, a model that can produce high-fidelity, free-viewpoint renderings of dynamic scenes with strong spatiotemporal consistency is required. Motivated by recent progress in 4D generation, we adopt TrajectoryCrafter~\cite{yu2025trajectorycrafter}, a pretrained video generative model capable of synthesizing high-fidelity videos along arbitrary camera trajectories from a single-view input video, as our 4D content generator.

Specifically, we first apply a monocular depth estimator~\cite{hu2025depthcrafter} to estimate the depth maps $\bm{D}^s = \{\bm{D}^s_i\}_{i=1}^n \in \mathbb{R}^{n \times h \times w}$ of the input video $\bm{V}^s$.
Subsequently, we lift $\bm{V}^s$ into a set of dynamic point clouds $\bm{P}=\{\bm{P}_i\}_{i=1}^n$ using inverse perspective projection $\bm{\Phi}^{-1}$, which is formulated as follows:
\begin{equation}
    \bm{P}_i=\Phi^{-1}([\bm{I}_i^s,\bm{D}_i^s],\bm{K}),
\end{equation}
where $\bm{K}\in\mathbb{R}^{3\times 3}$ is the camera intrinsic matrix. 
Using the estimated dynamic point clouds, we synthesize novel-view renderings $\bm{I}^r = \{\bm{I}^r_i\}_{i=1}^n$ by projecting $\bm{P}$ onto the target camera trajectory $\bm{T}^r = \{\bm{T}^r_i\}_{i=1}^n \in \mathbb{R}^{n \times 4 \times 4}$ using the following equation:
\begin{equation}
    \bm{I^r_i}=\Phi\left( \bm{T^r_i} \cdot \bm{P_i}, \bm{K} \right),
\end{equation}
where $\bm{\Phi}$ denotes the projection operation. 
The rendered results $\bm{I}^r$ are then used as conditioning inputs to the video diffusion model of TrajectoryCrafter, guiding the generation of high-fidelity novel-view videos.

Next, we prepare the raw monaural audio signal required for subsequent physics-based spatial audio simulation. To this end, we adopt MMAudio~\cite{cheng2024taming}, a transformer-based video-to-audio synthesis model, to generate monaural audio $\bm{A}^{m} \in \mathbb{R}^{1 \times T}$ that is temporally and semantically aligned with the input video $\bm{V}^s$. By decomposing the task of immersive 4D scene exploration into two sub-tasks—\textit{i.e.}, 4D scene generation and spatial audio synthesis—and leveraging pretrained expert models to extract the required visual and acoustic information, we can achieve greater flexibility and continually improve performance as upstream models evolve.

\subsection{Stage \uppercase\expandafter{\romannumeral2}: 3D Sound Source Localization and Tracking}\label{3D Sound Source Localization and Tracking}

With the dynamic point cloud $\bm{P}$ and its associated monaural audio $\bm{A}^m$, we proceed to localize and track the sound source in 3D space to facilitate physics-based spatial audio generation. However, directly tracking the sound source in a 4D environment remains highly challenging. Therefore, we first localize the sound source in the 2D pixel space and then back-project the estimated 2D coordinates into 3D space, thereby modeling the sound source trajectory within the generated dynamic scene.

To achieve this goal, existing 2D sound‑source localization methods~\cite{um2023audio, senocak2023sound, park2024can} typically require ground‑truth audio inputs and are often tailored to specific scenarios and input sizes. This dependence on “true” audio runs counter to Sonic4D’s very generation‑centric design. In contrast, given the rich commonsense knowledge and reasoning abilities of multimodal large language models (MLLMs), we propose to leverage GroundingGPT~\cite{li2024groundinggpt}, a highly generalizable MLLM focused on grounding tasks and trained on a variety of audio–visual datasets for sound source localization, as our 2D sound source localization component.

Concretely, as shown in Fig.~\ref{fig:pipeline}, we craft a textual prompt and feed the source video $\bm{V}^s$ frame by frame together with the prompt into GroundingGPT~\cite{li2024groundinggpt}. Assuming the grounding result for the $i$‑th frame $\bm{I}_i^s$ is represented by a bounding box $[x_0, y_0, x_1, y_1]$, we define the sound source pixel coordinates as the center of this bounding box, \textit{i.e.}:
\begin{equation}
    (u_i, v_i)
= \biggl(
\Bigl\lfloor \tfrac{x_0 + x_1}{2} \times W \Bigr\rfloor,\;
\Bigl\lfloor \tfrac{y_0 + y_1}{2} \times H \Bigr\rfloor
\biggr).
\end{equation}
where $W$ and $H$ are the width and height of $\bm{V}^s$. Let $d_i = \bm{D}(u_i, v_i)$ denote the depth at pixel $(u_i, v_i)$. Then the 3D point of the sound source in camera coordinate system corresponding to frame $i$ is given by
\begin{equation}
    \bm X_i \;=\; \pi^{-1}\bigl((u_i, v_i),\,d_i\bigr).
\end{equation}
Here, $\pi^{-1}$ denotes the back‑projection operator that maps a pixel coordinate and its depth into a 3D point.  Considering the sporadic errors introduced by monocular depth estimation at individual pixels, we compute the 3D coordinates of an \(r\times r\) pixel patch surrounding \((u_i,v_i)\) and average them to obtain the sound‐source’s 3D location as shown in Figure \ref{fig:stage2}:
\begin{equation}
    \bar{\bm{X}}_i
\;=\;
\frac{1}{\lvert \mathcal{P}^r(u_i,v_i)\rvert}
\sum_{(u,v) \in \mathcal{P}^r(u_i,v_i)}
\pi^{-1}\bigl((u,v),\,\bm{D}(u,v)\bigr).
\end{equation}
where $\mathcal{P}^r(u_i,v_i)$ denotes the set of all pixels $(u,v)$ such that $\lvert u - u_i\rvert \le (r-1)/2$ and $\lvert v - v_i\rvert \le (r-1)/2$.
Nonetheless, due to uncertainty in the bounding‐box predictions, the back‐projected trajectory \(\{\bar{\mathbf X}_i\}\) may still exhibit significant fluctuations. To mitigate this, we apply DBSCAN~\cite{ester1996density} (Density‐Based Spatial Clustering of Applications with Noise), a clustering‐based outlier detection algorithm that groups dense regions while marking sparse points as noise, to filter out spurious estimates.
Points marked as noise are then temporally filled in via linear interpolation between the nearest non–noise neighbours to produce a smooth trajectory $\textbf{Traj}_{src}=\{\bm{X}_i\}_{i=1}^n$.

In addition to the sound source's trajectory, the receiver trajectory is also required for acoustic simulation. We use the user‑specified camera trajectory as the receiver trajectory. For each frame \(i\), the camera pose is defined as
\begin{equation}
\bm T^r_i
=
\begin{bmatrix}
\bm R^r_i & \bm t^r_i \\[0.3em]
\bm 0^\mathsf{T} & 1
\end{bmatrix},
\end{equation}
where \(\bm R^r_i\in\mathbb{R}^{3\times3}\) is the rotation matrix and \(\bm t^r_i\in\mathbb{R}^3\) is the translation vector. The receiver’s 3D position is therefore
$\bm r_i \;=\;\bm t^r_i,$and the full receiver trajectory is $\textbf{Traj}_{rsv} \;=\;\{\bm r_i\}_{i=1}^n.$

\subsection{Stage \uppercase\expandafter{\romannumeral3}: Physics-Driven Spatial Audio Synthesis}\label{Physics-Driven Spatial Audio Synthesis}
Given the sound source's trajectory \(\textbf{Traj}_{src}\), the receiver trajectory \(\textbf{Traj}_{rsv}\), and the monaural audio signal \(\bm A^m\), we perform a physics‑based spatial audio simulation.

A central element of our simulation is the computation of Room Impulse Responses (RIRs) via the Image Source Method (ISM)~\cite{allen1979image}. Building on the Image‐Source Method for static RIRs, we apply gpuRIR~\cite{diaz2021gpurir} to simulate dynamic sources and receivers by segmenting the trajectories into short intervals over which they are assumed stationary. Concretely, let the total signal length be \(N\) samples and split it into \(M\) segments at sample indices
\[
0 = n_0 < n_1 < \cdots < n_M = N,
\]
For each segment \(i\in[1,M]\), let \(\mathbf x^s_i = \textbf{Traj}_{src}[i]\) denote the position of the sound source and \(\mathbf x^r_i = \textbf{Traj}_{rsv}[i]\) denote the position of the receiver. We simulate binaural microphone placement by offsetting from \(\mathbf x^r_i\) in the direction parallel to the camera’s rendering plane, yielding the left and right microphone positions \(\mathbf x^r_{i,L}\) and \(\mathbf x^r_{i,R}\). We compute the room impulse responses at the left and right microphones using the Image Source Method (ISM):
\begin{align}
h_{i,L}(\tau) = \mathrm{ISM}\bigl(\mathbf x^s_i,\;\mathbf x^r_{i,L}\bigr), \\
h_{i,R}(\tau) = \mathrm{ISM}\bigl(\mathbf x^s_i,\;\mathbf x^r_{i,R}\bigr).
\end{align}
Let the mono audio block for segment \(i\) be
\begin{equation}
b_i[k] = \bm A^m[n_i + k], 
\quad
0 \le k < n_{i+1}-n_i.
\end{equation}
Then the left and right binaural signals are obtained by block‐wise convolution:
\begin{align}
y_L[n] &= \sum_{i=0}^{M-1} \bigl(b_i * h_{i,L}\bigr)[\,n - n_i\,],\\
y_R[n] &= \sum_{i=0}^{M-1} \bigl(b_i * h_{i,R}\bigr)[\,n - n_i\,],
\end{align}
where \(h_{i,L}\) and \(h_{i,R}\) are the left‐ and right‐ear impulse responses in \(h_i\). Putting it all together, the final binaural signal is
\begin{equation}
\bm{A}^b = \bigl(y_L[n],\,y_R[n]\bigr).
\end{equation}
To ensure the output audio is in a valid range for playback, we normalize the signal to the range \([-1, 1]\):
\begin{equation}
\bm{A}^b \leftarrow \frac{\bm{A}^b}{\max\left(\left|\bm{A}^b\right|\right)}.
\end{equation}

Compared with learning-based methods for spatial audio generation, our physics-based audio rendering ensures physical plausibility by explicitly modeling sound propagation using room impulse responses, resulting in more realistic spatial cues such as directionality and reverberation. 

By decomposing the task into three modular stages, our framework achieves high flexibility and extensibility. Each stage leverages either pretrained models or physical priors, allowing Sonic4D to naturally scale with future advances. This decoupled design further enables plug-and-play upgrades, making Sonic4D applicable to a wide range of immersive content creation scenarios.
\section{Experiments}

\subsection{Experimental Settings}
Since there is no existing benchmark for spatial audio in 4D scene exploration, we use two complementary evaluations.

\subsubsection{SELD Evaluation.}
Inspired by \cite{shimada2024savgbench}, We adopt a pretrained Sound Event Localization and Detection(SELD) model~\cite{Diaz-Guerra2024}, which extracts a Multi-ACCDOA representation from 5s of video+stereo audio~\cite{shimada2022multi}. Following~\cite{Diaz-Guerra2024}, we evaluate on a curated subset of STARSS23~\cite{shimada2023starss23}, selecting perspective-view clips of musical instruments to avoid mosaic artifacts that may interfere with generative models. 

We compare our method with the state-of-the-art ViSAGe model~\cite{kim2025visage}. As ViSAGe requires additional source location inputs (elevation $\theta$ and azimuth $\phi$), we use the global mean elevation from its training dataset (YT-Ambigen), and for azimuth, we utilize the ground-truth annotations provided in STARSS23. Following the approach in~\cite{shimada2024savgbench}, we convert the ViSAGe-generated FOA audio into stereo format by first rotating the audio according to the azimuth, and then applying a simple transformation: $left = W + Y$ and $right = W - Y$, where $W$ is the omnidirectional component and $Y$ is the first-order horizontal (left-right) component of the FOA audio.

To quantify spatial audio performance, we employ the joint localization and detection metrics from the DCASE SELD challenge~\cite{Diaz-Guerra2024, shimada2024savgbench}. Specifically, we report:
\begin{itemize}
  \item \textbf{Location-aware F-score at 20° (F$_{20^\circ}$)}, which counts a detection as correct only if the class matches, the DoA angular error is within 20°, and the relative distance error is less than 1.0.
  \item \textbf{Directions of Arrival Angular Error ($\Delta_{\phi}$)}, the mean azimuth angular error. (Due to the SELD model, we only measure azimuth here.)
  \item \textbf{Distance Error ($\Delta_{d}$)} and \textbf{Relative Distance Error ($\Delta_{rd}$)}, measuring absolute and normalized distance estimation errors.
\end{itemize}
These metrics together capture both detection and localization quality in our spatial audio evaluations.

\subsubsection{User Study.}
We choose MMAudio~\cite{cheng2024taming} as the baseline method, whose generated audio is monaural and contains no spatialization. To determine whether our generated spatial audio conveys a sense of space, we paired the same viewpoint‑specific video with either the spatialized stereo audio or the non‑spatialized mono audio for comparison. Participants were first asked, for each video pair (identical visuals but different audio), to choose which audio track felt more spatial. Next, they rated both audio tracks on two criteria from 1 to 5:
\begin{itemize}
  \item \textbf{Spatial Localization Accuracy:} 
    Measures how faithfully the interaural level and time‐difference cues recreate the true 3D source position.  
    Participants rate from 1 (“no sense of source direction”) to 5 (“precise and stable perception of source location”).

  \item \textbf{Audio–Visual Spatial Alignment:} 
    Assesses how consistently the evolving audio cues (e.g.\ changes in loudness or binaural disparity) track the on‐screen motion of the subject or camera.  
    Participants rate from 1 (“audio movement conflicts with visuals”) to 5 (“audio shifts perfectly match visual motion”).
\end{itemize}
To more comprehensively demonstrate our model’s capabilities, we categorized the videos based on custom camera trajectories into \textbf{static} and \textbf{dynamic} viewpoints. Static viewpoints indicate using the original camera view unchanged or only modest transformations of that original view, with the viewpoint remaining fixed throughout; Dynamic viewpoints include translational camera movements, arc‑shaped camera trajectories, and push‑in/pull‑back motions.

\begin{figure*}[!htbp]
    \centering
    \begin{subfigure}{0.94\linewidth}
        \centering
        \includegraphics[width=\linewidth]{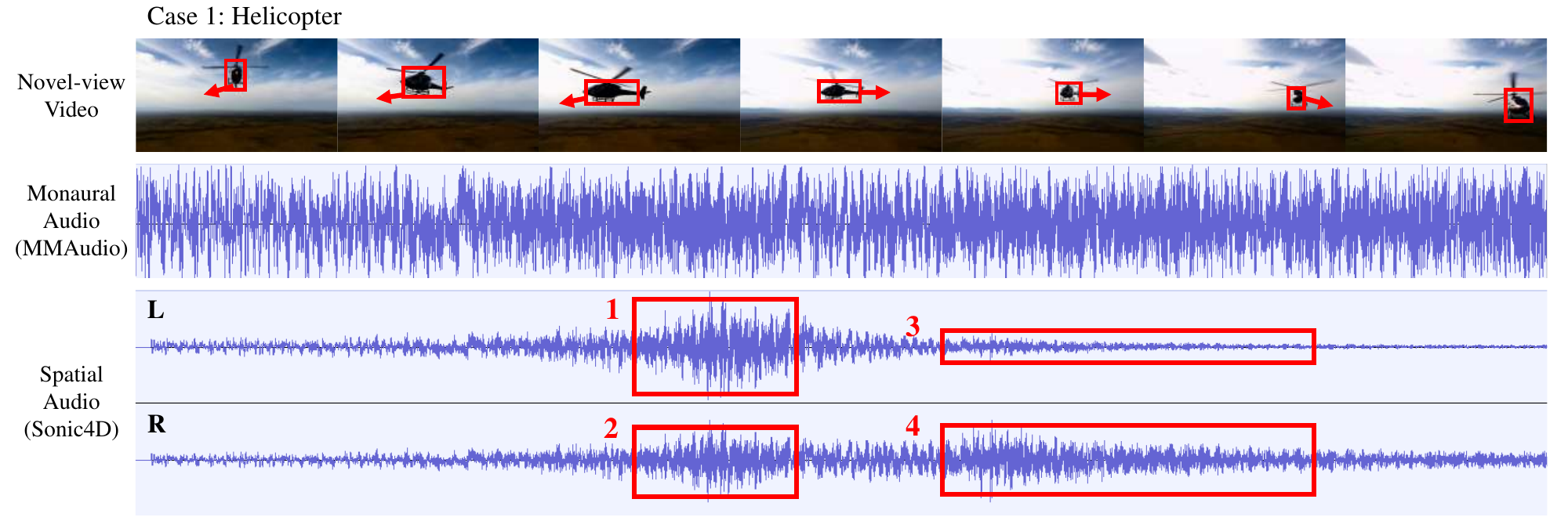}

    \end{subfigure}

    \vspace{0.01em} 

    \begin{subfigure}{0.94\linewidth}
        \centering
        \includegraphics[width=\linewidth]{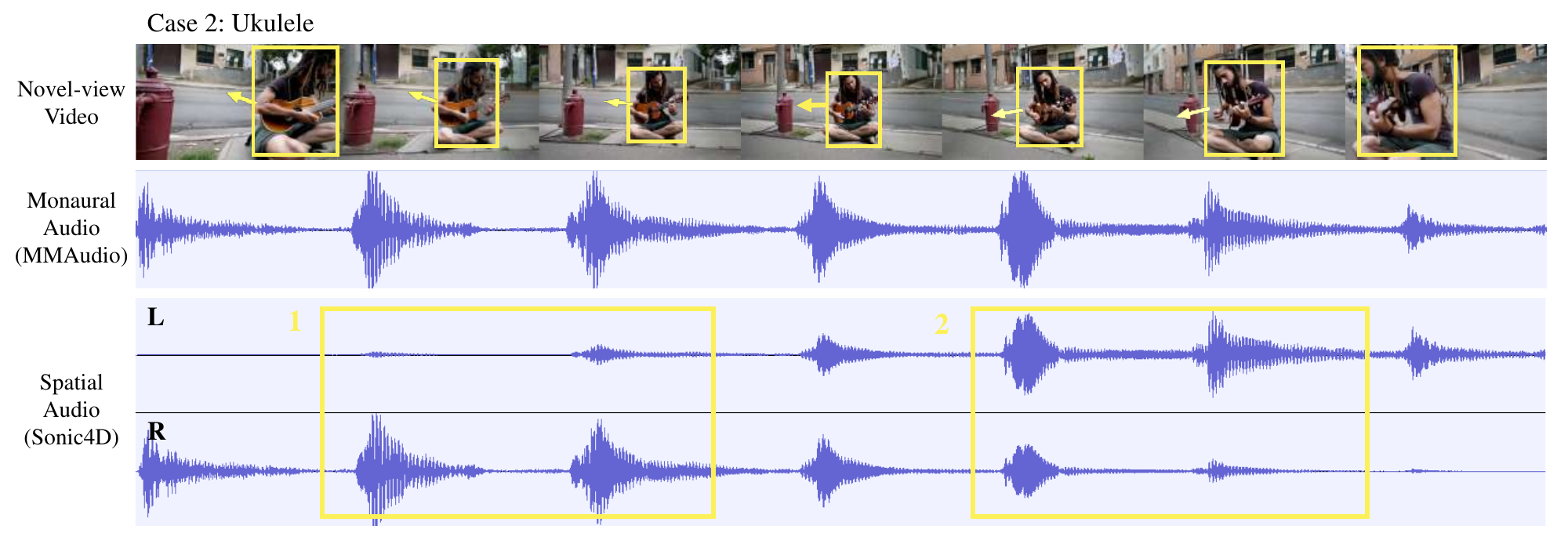}

    \end{subfigure}

    \vspace{0.01em}

    \begin{subfigure}{0.94\linewidth}
        \centering
        \includegraphics[width=\linewidth]{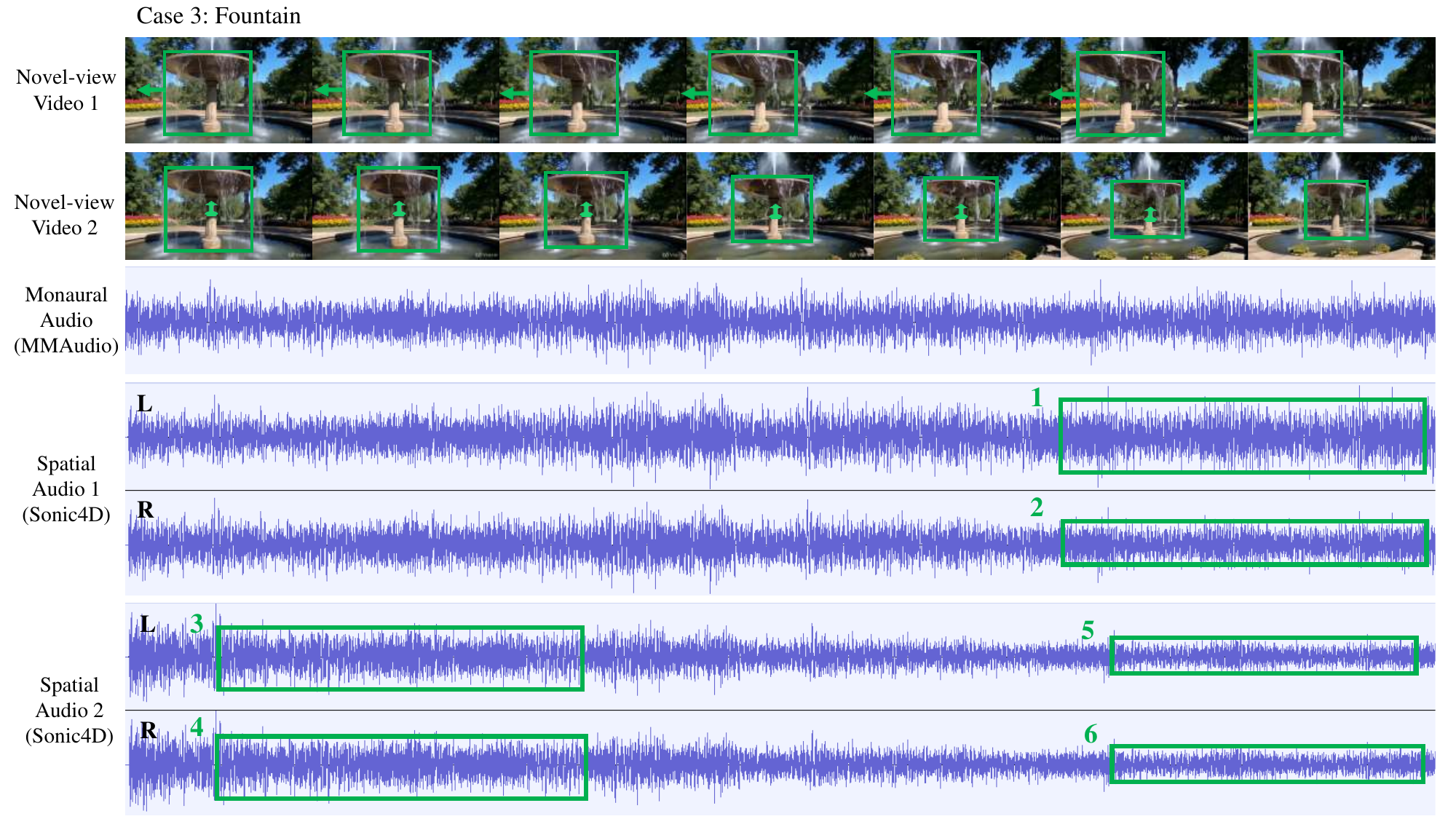}

    \end{subfigure}

    \caption{Qualitative results across different scenarios. We present comparisons of the spatial audio generated by Sonic4D conditioning on various camera trajectories, including static camera viewpoints, camera circling around the subject, rightward panning, and pulling out. These examples demonstrate the temporal and spatial alignment between the generated spatial audio and the motion of the visual subject.}
    \label{fig:qualitative_all}
\end{figure*}

\subsection{Quantitative Comparisons}
\subsubsection{SELD Evaluation.}
Table~\ref{tab:seld-metrics} summarizes the SELD evaluation results on our curated STARSS23 subset. Our method attains a relatively higher F-score (\textbf{4.0\%}) than ViSAGe, while reducing DOA angular error to \textbf{18.6\textdegree} and distance errors to \textbf{75.20} (absolute) and \textbf{0.44} (relative). These results demonstrate that Sonic4D produces more accurate spatial audio cues than purely learning-based methods, highlighting its ability to capture spatial information in a zero-shot setting.
\begin{table}[h]
    \centering
    \begin{tabular}{lcccc}
        \toprule
        \textbf{Method} & \textbf{F$_{20^\circ}$ $\uparrow$} & \textbf{$\Delta_{\phi}$(°) $\downarrow$} & \textbf{$\Delta_{d}$$\downarrow$} & \textbf{$\Delta_{rd}$ $\downarrow$} \\
        \midrule
        Ground truth & 6.5\%   & 7.2   & 23.67  & 0.15 \\
        ViSAGe      & 0.4\%   & 31.5  & 101.18 & 0.49 \\
        \midrule
        \textbf{Sonic4D}(\textit{ours}) & \textbf{4.0\%} & \textbf{18.6} & \textbf{75.20} & \textbf{0.44} \\
        \bottomrule
    \end{tabular}
    \caption{SELD evaluation results on the curated STARSS23 subset for Sonic4D vs. ViSAGe~\cite{kim2025visage}. }
    \label{tab:seld-metrics}
\end{table}
\subsubsection{User Study.}
We present the results of user study evaluation comparing the spatial audio rendered by our method (Sonic4D) against the non-spatialized mono audio from MMAudio~\cite{cheng2024taming}. As shown in Table~\ref{tab:subjective-results}, Sonic4D achieved a strong overall preference rate of \textbf{89.03\%}, indicating that the spatial audio generated by Sonic4D exhibits a significant perceptual difference in spatiality compared to the original mono audio, thus demonstrating the effectiveness of our model. In terms of Mean Opinion Scores (MOS), our method outperforms the baseline in both evaluation dimensions. For Spatial Localization Accuracy, Sonic4D achieved a mean score of \textbf{4.013}, compared to MMAudio~\cite{cheng2024taming}’s \textbf{2.322}. This suggests that participants were clearly able to perceive accurate and stable sound source positions when listening to our spatialized audio. A similar advantage is observed for Audio–Visual Spatial Alignment, where Sonic4D scored \textbf{3.977} versus \textbf{2.418} for MMAudio~\cite{cheng2024taming}, indicating a much stronger alignment between auditory motion cues and on-screen visual dynamics.

\begin{table}[ht]
\centering
\begin{tabular}{lccc}
\toprule
\textbf{Metric} & \textbf{All} & \textbf{Static} & \textbf{Dynamic} \\
\midrule
\multicolumn{4}{l}{\textit{Preference (\%)}} \\
\quad MMAudio             & 10.97\%           & 12.41\%            & 9.90\%             \\
\quad \textbf{Sonic4D} (\textit{ours}) & \textbf{89.03\%}  & \textbf{87.59\%}   & \textbf{90.10\%}   \\
\midrule
\multicolumn{4}{l}{\textit{MOS-SLA}} \\
\quad MMAudio              & 2.322             & 2.357              & 2.296              \\
\quad \textbf{Sonic4D} (\textit{ours}) & \textbf{4.013}    & \textbf{3.986}     & \textbf{4.033}     \\
\midrule
\multicolumn{4}{l}{\textit{MOS-AVSA}} \\
\quad MMAudio              & 2.418             & 2.493              & 2.363              \\
\quad \textbf{Sonic4D} (\textit{ours}) & \textbf{3.977}    & \textbf{3.980}     & \textbf{3.975}     \\
\bottomrule
\end{tabular}
\vspace{2mm}
\caption{Subjective evaluation results for Sonic4D vs. MMAudio~\cite{cheng2024taming}.}
\label{tab:subjective-results}
\end{table}

\subsection{Qualitative Comparisons}

As shown in Fig.~\ref{fig:qualitative_all}, we present additional qualitative comparisons with the baseline method. We predefined three types of camera trajectories: static viewpoint (case 1), camera circling around the subject (case 2), rightward panning and pulling out (case 3). For each case, we show the rendered video from the novel viewpoint, annotated with the motion direction of the visual subject within the frame. We then present the original waveform along with the stereo spatial audio waveform generated by Sonic4D. We highly recommend using headphones to listen to the specific examples provided in our supplementary material.

In Case 1, as the helicopter moves left then right, the spatial audio waveform reflects this motion: early segments (\textbf{Regions 1} and \textbf{2}) show the left channel dominant, whereas later segments (\textbf{Regions 3} and \textbf{4}) show the right channel surpassing the left corresponding to the helicopter’s trajectory, aligning well with the expected spatial perception of the moving object. In case 2, the camera starts on the right side of a ukulele player and orbits to the left. The left-right disparity in \textbf{Regions 1} and \textbf{2} (Yellow), as well as the temporal progression within those regions, closely mirrors the trajectory, indicating strong spatial tracking in the generated audio. Case 3 demonstrates that the same source video can yield notably different spatial audio depending on the camera motion, validating the view-dependence of our model. In video 1, as the camera pans rightward, the fountain becomes closer to the left ear, leading to increased amplitude in the left channel (\textbf{Regions 1, 2}, Green). In video 2, the camera gradually pulls out, producing a fading splash sound that matches the increasing distance (\textbf{Regions 3, 4, 5, 6}, Green).

These qualitative results demonstrate that the spatial audio generated by Sonic4D aligns closely with the visual cues, reflecting both spatial position and dynamic changes in the scene.

\section{Conclusions}
In this paper, we present Sonic4D, a novel paradigm that enables spatial audio generation for immersive 4D scene exploration. Specifically, our method consists of three stages: \textbf{1) Dynamic Scene and Monaural Audio Generation:} we first employ pre-trained expert models to generate the 4D scene and its corresponding monaural audio, aiming to provide sufficient spatial and acoustic priors for subsequent spatial audio simulation. \textbf{2) 3D Sound-Source Localization and Tracking:} we further propose a pixel-level visual grounding strategy to estimate the sound source’s trajectory in 3D space, facilitating spatial alignment between audio and visual content in dynamic scenes. \textbf{3) Physics-Driven Spatial Audio Synthesis:} Based on the estimated sound source's trajectory, we synthesize plausible, viewpoint-adaptive spatial audio using physics-based room impulse response (RIR) simulation. Extensive experiments demonstrate that our proposed method generates realistic spatial audio that is consistent with the synthesized 4D scene in a training-free manner, significantly enhancing the user’s immersive experience.

\bibliography{aaai2026}

\end{document}